\documentclass{emulateapj}
\usepackage{graphicx}

\def\s4u0142{4U~0142+61~}
\def\a4u0142{AXP~4U~0142+61~}
\def\rxs{1RXS~J170849.0-400910~}
\def\rxsnos{1RXS~J170849.0-400910}
\def\kes1841{1E~1841-045}
\def\ctb2259{1E~2259+586}
\def\e1048{1E~1048.1-5937}
\def\CXOU{CXOU~J164710.2-455216}

\begin{document} 

\title{Long-Term Timing and Glitch Characteristics of Anomalous X-ray Pulsar \rxs} 

\author{Sinem \c{S}a\c{s}maz Mu\c{s}\altaffilmark{1},
Ersin G\"o\u{g}\"u\c{s}\altaffilmark{1}}

\altaffiltext{1}{Sabanc\i~University, Faculty of Engineering and Natural Sciences, Orhanl\i $-$ Tuzla, 34956 Istanbul Turkey}

\email{sinemsm@sabanciuniv.edu} 

%%%%%%%%%%%%%%%%%%%%%%%%%%%%%%%%%%%%%%%%%%%%%%%%%%%%%%%%%%%%%%%%%%%%%%%%%
%ABSTRACT
%%%%%%%%%%%%%%%%%%%%%%%%%%%%%%%%%%%%%%%%%%%%%%%%%%%%%%%%%%%%%%%%%%%%%%%%%

\begin{abstract}  

We present the results of our detailed timing studies of an anomalous 
X-ray pulsar, \rxsnos, using {\it Rossi X-ray Timing Explorer} ({\it RXTE}) 
observations spanning over $\sim$6 yr from 2005 until the end of 
{\it RXTE} mission. We constructed the long-term spin characteristics of 
the source and investigated time and energy dependence of pulse 
profile and pulsed count rates. We find that pulse profile and 
pulsed count rates in the 2$-$10 keV band do not show any significant 
variations in $\sim$6 yr. \rxs has been the most frequently 
glitching anomalous X-ray pulsar: three spin-up glitches and three candidate 
glitches were observed prior to 2005. Our extensive search for glitches 
later in the timeline resulted in no unambiguous glitches though we 
identified two glitch candidates (with $\Delta{\nu}/\nu$ $\sim$$10^{-6}$) 
in two data gaps: a strong candidate around MJD 55532 and another 
one around MJD 54819, which is slightly less robust. We discuss our 
results in the context of pulsar glitch models and expectancy of 
glitches within the vortex unpinning model.

\end{abstract} 


%%%%%%%%%%%%%%%%%%%%%%%%%%%%%%%%%%%%%%%%%%%%%%%%%%%%%%%%%%%%%%%%%%%%%%%%%
%INTRODUCTION
%%%%%%%%%%%%%%%%%%%%%%%%%%%%%%%%%%%%%%%%%%%%%%%%%%%%%%%%%%%%%%%%%%%%%%%%%
\section{Introduction} 

Glitches, sudden jumps in the rotation frequency of neutron stars, are the 
unique events that provide invaluable information on the internal structure 
of extremely compact stars. Originally detected from rotation powered 
neutron stars \citep[see e.g.,][]{richards69,radha69}, glitches are 
generically not associated to changes in the radiative behavior of the source. 
\citep[but see,][]{weltevrede11}. Therefore, the proposed glitch models involve 
dynamical variations in the neutron star interior instead of an external torque 
mechanism. The size of the glitch typically reflects the underlying internal 
dynamics of the neutron star: small-size glitches ($\Delta{\nu}$/$\nu$$\sim$10$^{-9}$, 
aka. Crab$-$like glitches) are explained by the decrease of the moment of inertia 
of the pulsar \citep{ruderman69,baym71} and large-size glitches 
($\Delta{\nu}$/$\nu$$\sim$10$^{-6}$, aka. Vela$-$like glitches) are described as 
the angular momentum transfer from inner crust neutron superfluid to the crust by 
the sudden unpinning of the vortices that are pinned to the inner crust nuclei 
\citep{anderson75,pines80}.

Anomalous X-ray Pulsars (AXPs) are slowly rotating (P $\sim$ 2$-$12 s) neutron 
stars with persistent emission being significantly in excess of their inferred 
rotational energy loss rate. So far, there has been no evidence of binary 
signature in AXPs. They are young systems ($\sim$10$^4$ yr) as inferred from 
their characteristic spin-down ages (P/2$\dot{P}$), and also supported by their 
location on the plane of Milky Way, and the association of at least five AXPs 
with their supernova remnants. Almost all AXPs emitted short duration, energetic 
bursts in X-rays (see, e.g., \citealt{gavriil02,kaspi03} and for a recent review 
\citealt{reaesposito11}). Their surface dipole magnetic field strengths inferred 
from their periods and spin-down rates are on the order of 10$^{14}$ $-$ 10$^{15}$ 
G, which is much higher than that of conventional magnetic field strengths of pulsars. 
The decay of their extremely strong magnetic fields is proposed as the source 
of energy for their persistent X-ray emission and burst activity 
\citep{thompson95,thompson96,thompson02}. Recently, observational evidence of 
dipole field decay was reported by \cite{dosso12}.

Glitch activity from an AXP was first seen in \rxs \citep{kaspi00}. Thanks to 
almost continuous spin monitoring of AXPs with {\it RXTE} for more than a 
decade, sudden spin frequency jumps have now been observed from six AXPs 
\citep[see, e.g.,][]{kaspi03,dosso03,woods04,morii05,israel07a,israel07b,
dib08,dib09,gavriil11}. Fractional glitch amplitudes ($\Delta{\nu}$/$\nu$) of 
these events range from 10$^{-8}$ to 10$^{-5}$ \citep{dib09,icdem12} and 
fractional postglitch change in spin-down rates ($\Delta{\dot{\nu}}$/$\dot{\nu}$) 
are between $-$0.1 and 1 \citep{kaspi03,dib09}.

Glitches from AXPs somehow resemble those from radio pulsars, but contain 
some peculiar distinctive features in their recovery behavior and associated 
radiative characteristics \citep{woods04,morii05,dib08,dib09,gavriil11}. AXP 
\ctb2259 went into an outburst in conjunction with a glitch \citep{kaspi03,woods04}. 
AXP \e1048 has shown X-ray burst correlated with a glitch event \citep{dib09}. 
During the burst active phase of \a4u0142 between 2006 and 2007, six short bursts 
and a glitch with a long recovery time were observed \citep{gavriil11}. 
AXP \kes1841 has exhibited bursts and glitches, but not coincidentally 
\citep{dib08,zhu10,kumarsafi10,lin11}. \cite{israel07b} reported a burst 
and an extremely large glitch ($\Delta{\nu}$/$\nu$ $\sim$ 6 $\times$ 10$^{-5}$) 
from \CXOU, but the possibility of such a glitch was ruled out by 
\cite{woods11}. However, the latter team point out that a glitch with the 
size of usual AXP glitches may indeed have occurred. \rxs has been the most 
frequently glitching AXP \citep{kaspi00,kaspi03,dosso03,israel07a,dib08}, 
but it has not shown any bursts or remarkable flux variability related to the 
glitch epochs.

It is still unclear whether glitches are always associated with radiative 
enhancements. Recently, \cite{pons12} suggested that in the context of the 
starquake model, glitches observed in the bright sources can be related to 
the radiative enhancements but due to the bright quiescent state of these 
sources and fast decay of the enhancements, these events can be observed 
as small changes in the luminosity or only detected in faint sources. 

\rxs is an AXP with a spin period of $\sim$11 s. After the discovery 
of its spin period \citep{sugizaki97}, it has been monitored with 
{\it RXTE} for $\sim$13.8 yr. Analyzing the first $\sim$1.4 yr 
of data \cite{israel99} and \cite{kaspi99} have concluded that the 
source is a stable rotator. The continued monitoring has been essential 
in detecting three unambiguous glitches and three glitch candidates 
without any significant pulse profile variations 
\citep{kaspi00,kaspi03,dosso03,israel07a,dib08}. There appears to be 
a correlation between intensity and spectral hardness: the X-ray 
spectrum gets softer(harder) while the X-ray flux decreases(increases), 
possibly in relation with glitches \citep{rea05,campana07,rea07,israel07a}. 
\cite{gotz07} reported the same correlation in the hard X-rays 
using {\it INTEGRAL}/ISGRI data. However, \cite{hartog08} claimed that 
they did not find the reported variability in their analysis. \cite{thompson02} 
proposed that external magnetic field can twist and untwist. Twisting 
and untwisting of the external magnetic field can lead to cracks and 
unpin the vortices for the glitches \citep{thompson96,dosso03}. Such 
twist/untwist of the magnetic field with a period of $\sim$5$-$10 yr 
has been suggested as an explanation for the observed correlations 
\citep{rea05,campana07}.

Here, we report on the analysis of long-term {\it RXTE} observations of \rxs 
spanning $\sim$6 yr. In \S\ref{sect:obs} we describe {\it RXTE} observations 
that we used in our analysis. We present long-term timing characteristics 
of the source in \S\ref{sect:timing}. In \S\ref{sect:pulseprofile} \& 
\S\ref{sect:pulsedrates} we constructed the pulse profiles, calculated 
pulsed count rates and examined their variability both in time and energy. 
We present the results of our extensive search for glitches in 
\S\ref{sect:glitches}. Finally, in \S\ref{sect:discuss} we discuss our 
results in the context of glitch models and expectancy of glitches in the 
vortex unpinning model.

%%%%%%%%%%%%%%%%%%%%%%%%%%%%%%%%%%%%%%%%%%%%%%%%%%%%%%%%%%%%%%%%%%%%%%%%%
%OBSERVATION AND DATA ANALYSIS
%%%%%%%%%%%%%%%%%%%%%%%%%%%%%%%%%%%%%%%%%%%%%%%%%%%%%%%%%%%%%%%%%%%%%%%%%
\section{{\it RXTE} Observations}
\label{sect:obs}

\begin{figure}[h]
\begin{center}
\vskip -0.5cm
\includegraphics[scale=0.6]{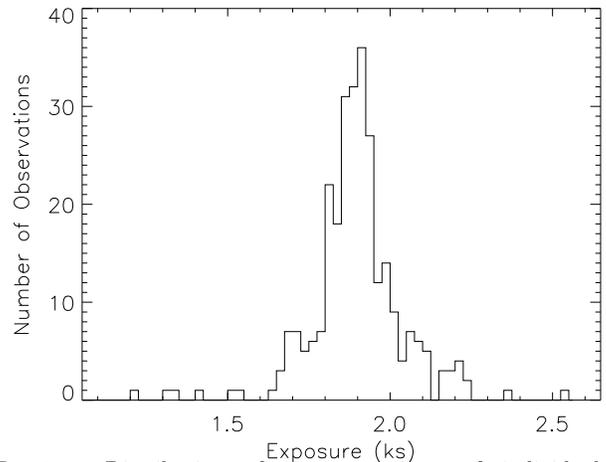}
\vskip -0.5cm
\caption{Distribution of exposure times of individual 
{\it RXTE}/PCA observations. The shortest observation 
with an exposure of 0.25 ks is excluded for clarity.}
\label{fig:exposurehist}
\end{center}
\end{figure}

\rxs has been almost regularly monitored with {\it RXTE} in 528 pointings 
since the beginning of 1998. Phase connected timing behavior of the source 
was investigated by \cite{dib08} using the {\it RXTE} data collected 
between 1998 January 12 and 2006 October 7, \cite{dosso03} using data from 
1998 January 13 to 2002 May 29, and \cite{israel07a} using from 2003 January 
5 to 2006 June 3. Here we analyzed {\it RXTE} data collected in 280 pointings 
between 2005 September 25 and 2011 November 17 with the Proportional Counter 
Array (PCA). Note that the first 49 pointings in our sample were also 
used by \cite{dib08}. We included them in order to maintain the continuity 
in the timing characteristics of \rxsnos. Exposure times of individual {\it RXTE} 
observations ranged between 0.25 ks (in one observation) and 2.5 ks, with a 
mean exposure time of 1.9 ks (see Figure~\ref{fig:exposurehist} for a 
distribution of exposure times). For our timing analysis, we used data collected 
with all operating Proportional Counter Units (PCUs) in GoodXenon mode that 
provides a fine time resolution of 1 $\mu$s.

%FIGURE 1
%%%%%%%%%%%%%%%%%%%%%%%%%%%%%%%%%%%%%%%%%%%%%%%%%%%

%%%%%%%%%%%%%%%%%%%%%%%%%%%%%%%%%%   TIMING   %%%%%%%%%%%%%%%%%%%%%%%%%%%%%%%%%%%%%%%

\section{Data Analysis And Results}

\subsection{Phase Coherent Timing}
\label{sect:timing}

\begin{figure*}[t]
\vspace{-1.0cm}
\begin{center}
\includegraphics[scale=0.7]{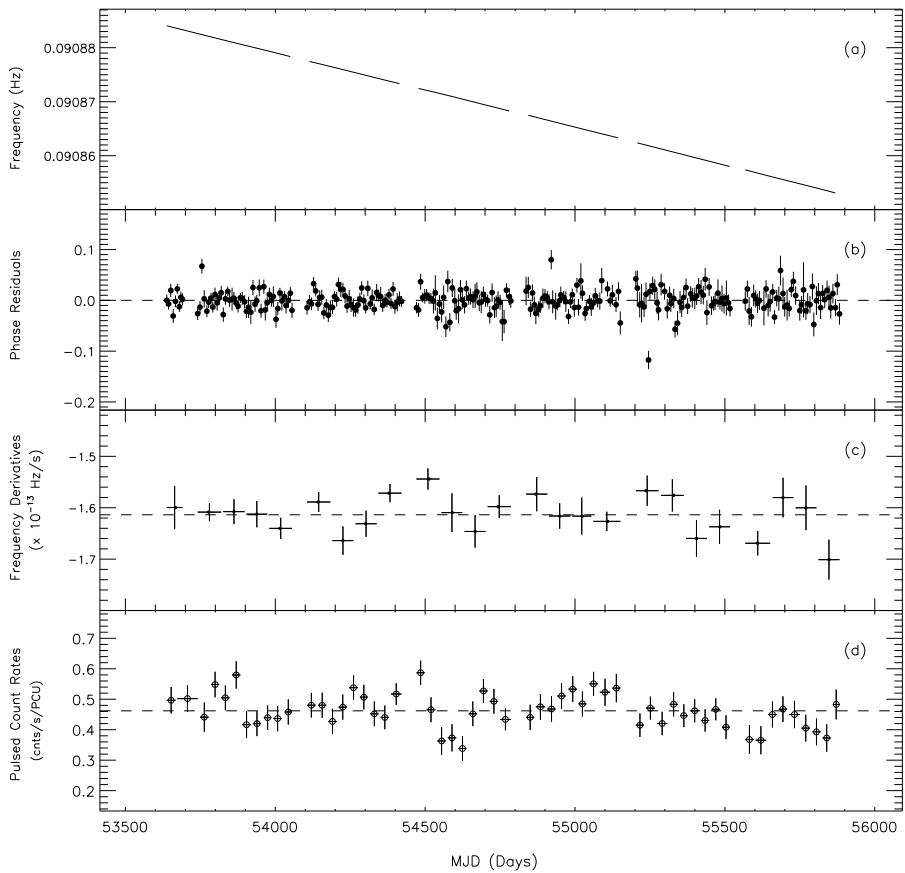}
\vspace{-0.5cm}
\caption{(a) Spin frequency evolution of \rxsnos. (b) Phase residuals 
after the subtraction of the pulse ephemeris given in Table 
\ref{tab:tablemain}. (c) Frequency derivatives obtained using 
$\sim$2.5 months long data segments. (d) Long term behavior of the 
rms pulsed count rates in the 2$-$10 keV band.}
\label{fig:spindervflux}
\end{center}
\end{figure*}

We selected events in the 2$-$6 keV energy range from the top Xenon layer 
of each PCU in order to maximize the signal-to-noise ratio, as done 
also by \cite{dib08}. All event arrival times were converted to the solar 
system barycenter and binned into light curves of 31.25 ms time resolution. 
We inspected each light curve for bursts and discarded the time intervals 
with the instrumental rate jumps. We merged observations together if the 
time gap between them was less than 0.1 days. The first set of observations 
(i.e., segment 0 in Table~\ref{tab:tablemain}) which includes 49 observations 
from \cite{dib08} were folded initially with the spin ephemeris given by 
\cite{dib08} and later by maintaining the phase coherence. We then cross-correlated 
the folded pulse profiles with a high signal-to-noise template pulse profile 
generated from a subset of observations and determined the phase shifts of 
observations with respect to the template. We fitted phase shifts with
\begin{equation}
\phi(t) = \phi_{0}(t_{0})+\nu_{0}(t-t_{0})+ \frac{1}{2}\dot{\nu_{0}}(t-t_{0})^{2} + ...,
\label{eq:taylorphase}
\end{equation}
whose coefficients yield the spin frequency, and its higher order time 
derivatives, if required. In Table~\ref{tab:tablemain} we list the best fit 
spin frequency and frequency derivatives to the specified time intervals, 
obtained using also listed number of time of arrivals (TOAs).
In Figure~\ref{fig:spindervflux} (a) we present the spin frequency evolution 
of \rxsnos, and in (b) phase residuals after subtraction of the best fit 
phase model given in Table~\ref{tab:tablemain}. We obtained frequency 
derivatives by fitting a second order polynomial to the sub-intervals of 
about 2.5 months long data and present them in Figure~\ref{fig:spindervflux} (c).

%TABLE 1
%%%%%%%%%%%%%%%%%%%%%%%%%%%%%%%%%%%%%%%%%%%%%%%%%%%%%%%%%%%%
\begin{deluxetable*}{lcccccc}[p]
\tablewidth{0pt}
\tabletypesize{\scriptsize}
%\rotate
\tablecaption{Pulse Ephemeris of \rxs$^{a}$
\label{tab:tablemain}}
\tablehead{
\colhead{Parameter}&
\colhead{Segment 0}&
\colhead{Segment 1}&
\colhead{Segment 2}&
\colhead{Segment 3}&
\colhead{Segment 4}&
\colhead{Segment 5}
}
\startdata
Range (MJD)  & 53638 $-$ 54056 & 54106 $-$ 54421 & 54471 $-$ 54786 & 54837 $-$ 55151 & 55203$-$ 55517 & 55568 $-$ 55882 \\
Epoch (MJD) & 53635.6772 & 54106.040 & 54471.050 & 54836.804 & 55202.849 & 55567.977\\
Number of TOAs   & 55 & 46 & 46 & 43 & 45 & 44 \\
$\nu$ (Hz)  & 0.090884080(5)& 0.090877558(5) & 0.090872536(5)& 0.090867590(3) & 0.090862448(1) & 0.090857386(7) \\
$\dot{\nu}$  ($10^{-13}$ Hz s$^{-1}$)  & $-$1.55(2) & $-$1.50(1)& $-$1.43(2) & $-$1.642(5)& $-$1.641(1) & $-$1.73(2) \\
$\ddot{\nu}$ ($10^{-22}$ Hz s$^{-2}$)  & $-$18(4) & $-$14(2)& $-$23(3)& 1.9(4)& $-$ & 19(3) \\
$d^{3}\nu/dt^{3}$ ($10^{-28}$ Hz s$^{-3}$)  & 2.9(5) &  0.9(2) & 1.3(2)& $-$ & $-$ & -1.41(3) \\
$d^{4}\nu/dt^{4}$ ($10^{-35}$ Hz s$^{-4}$) & $-$1.7(2) & $-$ & $-$ & $-$ & $-$ & $-$ \\
rms (phase) & 0.0174 & 0.0145 & 0.0200 & 0.0212 & 0.0265 & 0.0203
\enddata
\tablenotetext{a}{Values in parenthesis are the uncertainties in 
the last digits of their associated measurements}
\end{deluxetable*}

%FIGURE 2
%%%%%%%%%%%%%%%%%%%%%%%%%%%%%%%%%%%%%%%%%%%%%%%

%%%%%%%%%%%%%%%%%%%%%%%%%%%%%%  PULSE PROFILE  %%%%%%%%%%%%%%%%%%%%%%%%%%%%

\subsection{Pulse Profile Evolution}
\label{sect:pulseprofile}

\begin{figure*}[p]
\vspace{-0.2cm}
\begin{center}
\includegraphics[scale=0.8]{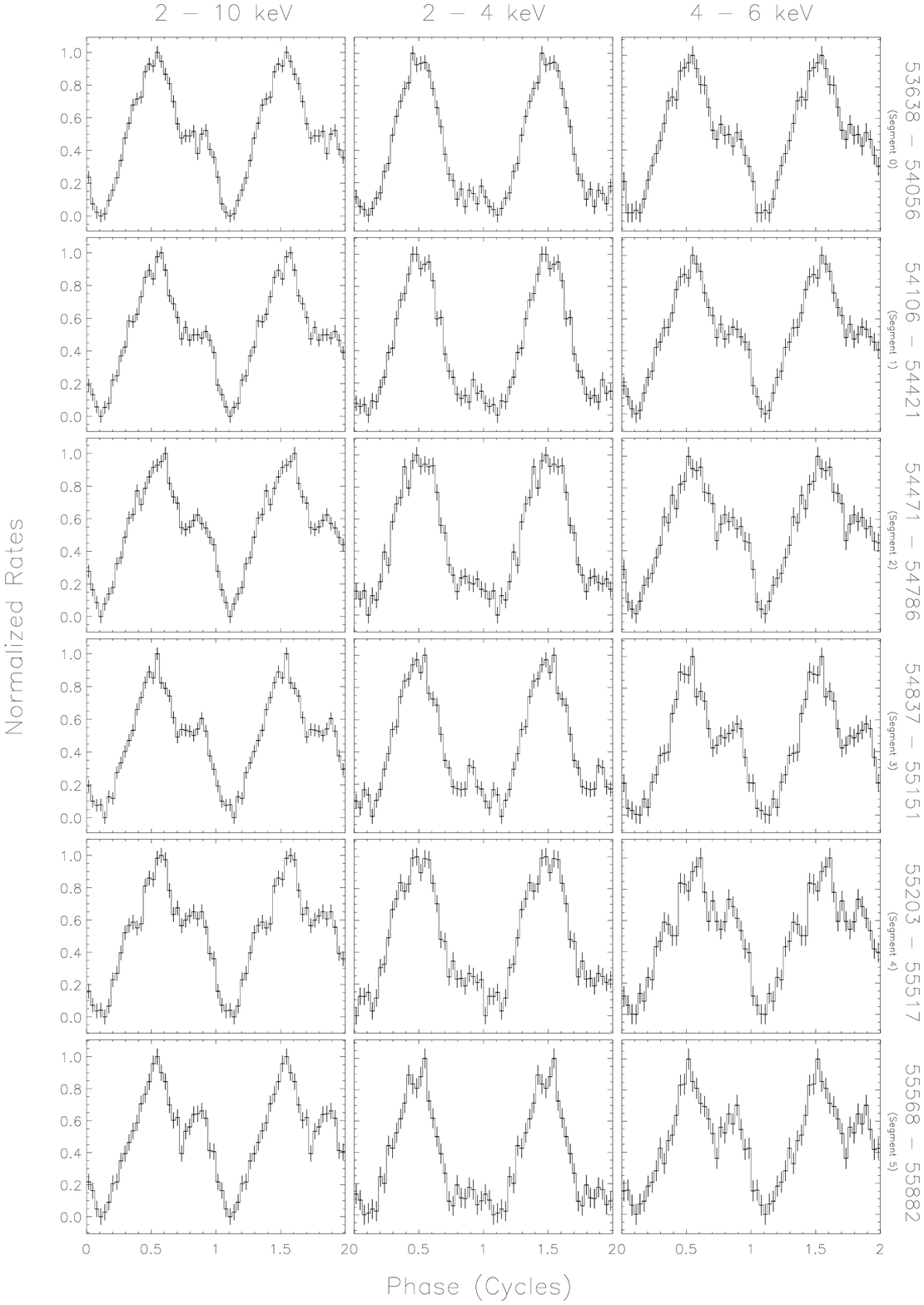}
\vspace{0.5cm}
\caption{Pulse profile history of \rxs in the energy bands 2$-$10, 
2$-$4 and 4$-$6 keV. The labels on the right are the corresponding 
time intervals of accumulated data.}
\label{fig:1708prof1}
\end{center}
\end{figure*}

\begin{figure*}[p]
\vspace{0.0in}
\begin{center}
\includegraphics[scale=0.8]{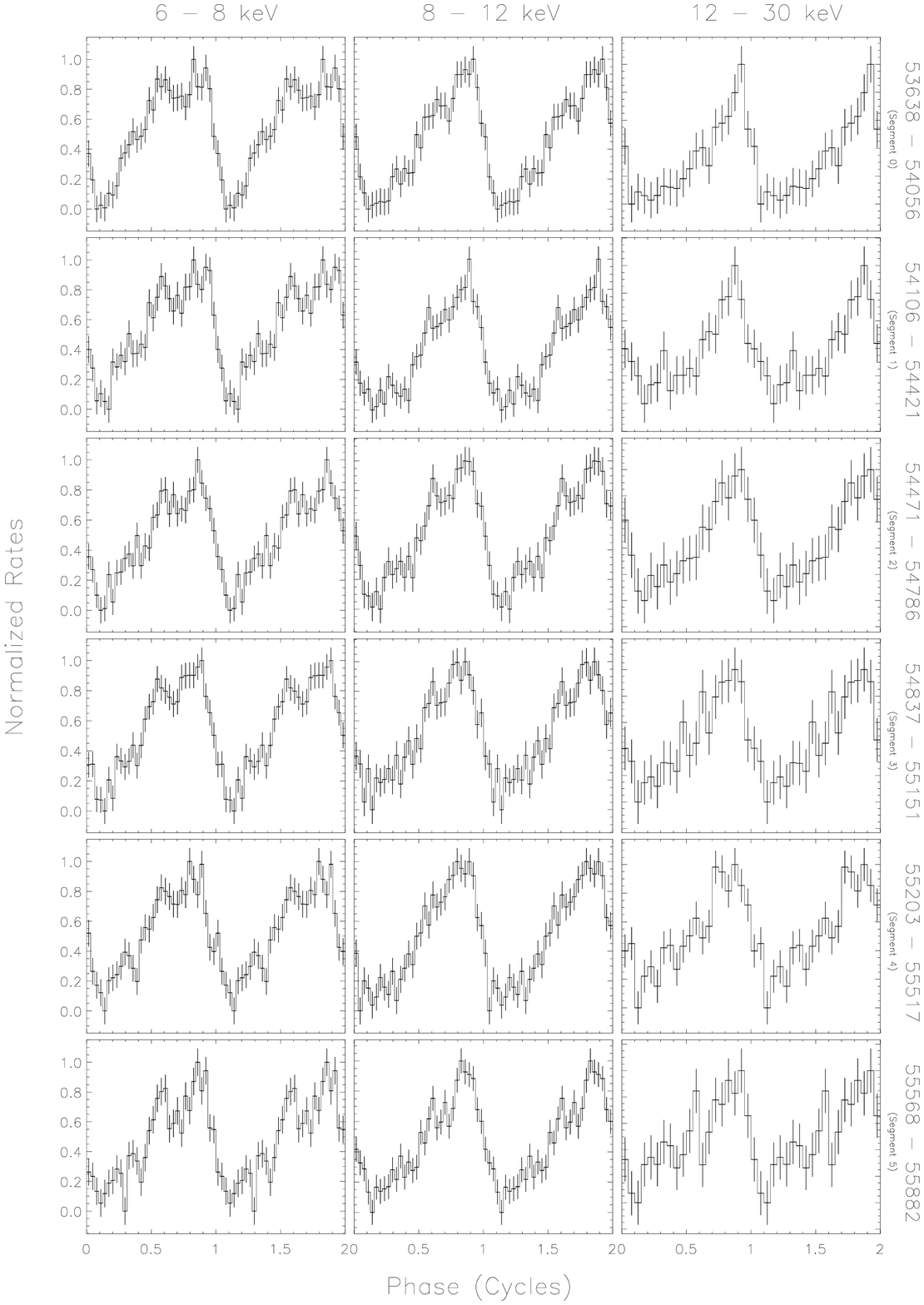}
\vspace{0.5cm}
\caption{Pulse profile history of \rxs in the energy bands 6$-$8, 
8$-$12 and 12$-$30 keV. The labels on the right are the 
corresponding time intervals of accumulated data.The 12$-$30 keV 
profiles are plotted with 20 phase bins due to lower count rate 
in this energy band.}
\label{fig:1708prof2}
\end{center}
\end{figure*}

We investigated long term pulse profile evolution of the source both 
in energy and time. For the pulse profile analysis, we excluded data 
collected with PCU0 and the data of PCU1 for the observations after 
2006 December 25 due to the loss of their propane layers (therefore, 
having elevated background levels). We obtained the pulse profiles with 
32 phase bins by folding the data in six energy bands with the appropriate 
phase connected spin ephemeris given in Table~\ref{tab:tablemain}. 
The energy intervals investigated are 2$-$10 keV, 2$-$4 keV, 4$-$6 keV, 
6$-$8 keV, 8$-$12 keV and 12$-$30 keV. In order to account for the 
different number of operating PCUs, we normalized the rates of each bin 
with the number of active PCUs. Finally, we subtracted the 
DC level and divided by the maximum rate of each profile. 
In Figures~\ref{fig:1708prof1} and \ref{fig:1708prof2}, we present the 
normalized pulse profiles for the six segments given in 
Table~\ref{tab:tablemain} in six energy bands and their evolution in time. 

%FIGURE 3
%%%%%%%%%%%%%%%%%%%%%%%%%%%%%%%%%%%%%%%%%%%%%

%FIGURE 4
%%%%%%%%%%%%%%%%%%%%%%%%%%%%%%%%%%%%%%%%%%%%%

The 2$-$10 keV pulse profiles of \rxs are characterized by a broad 
structure formed by the superposition of two features: the main 
peak near the pulse phase, $\phi\sim$0.55 and a weaker shoulder 
around $\phi\sim$ 0.85. Pulse profiles of the two lowest energy 
bands exhibit an additional shoulder (near phase $\sim$0.35) in 
the 55203 $-$ 55516 epoch (Segment 4), which is not clearly seen 
in any other epochs. Pulse profiles in the 2$-$4 keV band consist 
of the main peak in all epochs, while the shoulder feature 
($\phi\sim$ 0.85) is either weak or non-existent. The shoulder appears 
in the 4$-$6 keV band, and becomes more dominant above 6 keV. Pulse 
profiles above 8 keV contain only the shoulder feature. Note the fact 
that the duty cycle of the pulse profiles drops with increasing energy. 
The dominance of the secondary peak (shoulder) with the increase in 
photon energy was also reported in \cite{hartog08} by using {\it INTEGRAL}, 
{\it XMM$-$Newton} and earlier {\it RXTE} observations. 

\begin{figure}[h]
\vspace{0.0cm}
\begin{center}
\includegraphics[width=1.0\columnwidth]{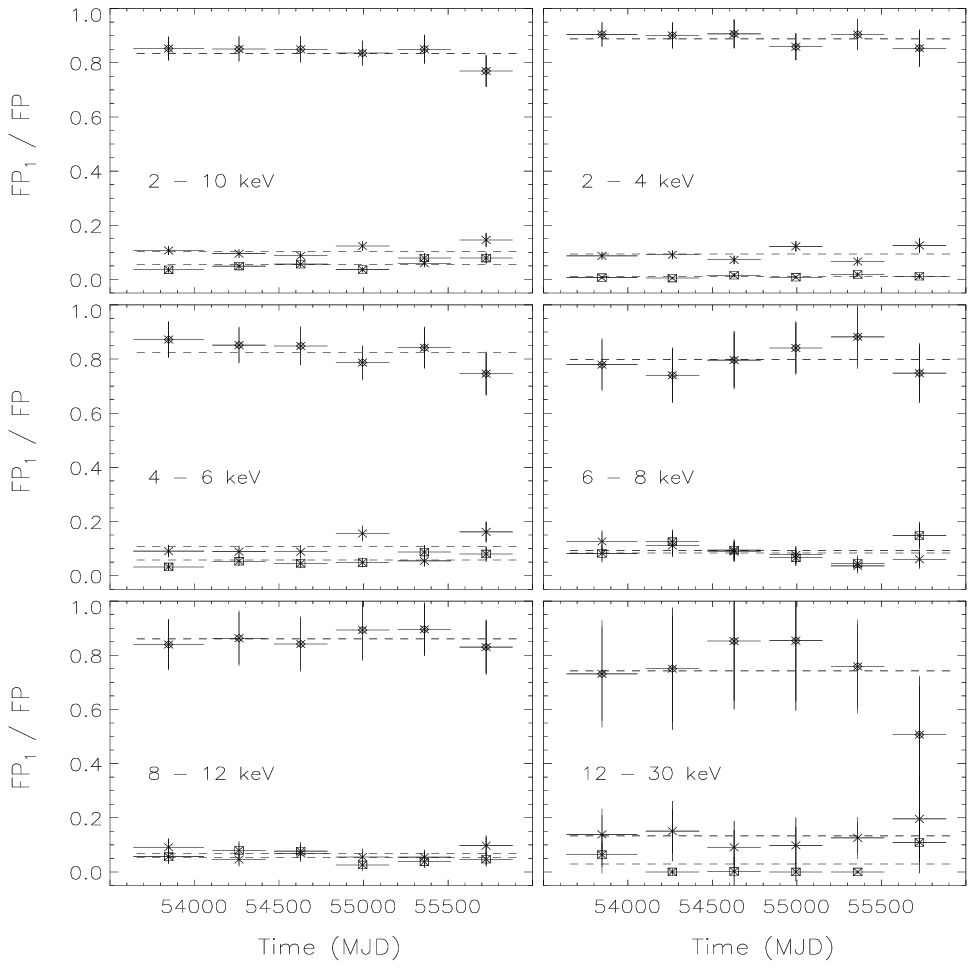}
\caption{Time evolution of the normalized Fourier harmonic powers 
in the first three harmonics. Dashed lines represent the averaged 
power of the related harmonic in all segments. The energy intervals 
in which the powers are calculated are displayed inside the panels.}
\label{fig:harmpow}
\end{center}
\end{figure}

We calculated the Fourier Powers (FPs) for a quantitative measure of 
the pulse profile variations. First we computed the Fourier transform of 
each profile and calculated the powers in the first six harmonics as 
FP$_{\rm k}$ = 2(a$_{\rm k}^2$~+~b$_{\rm k}^2$)/($\sigma^2_{\rm a_{\rm k}}~+~\sigma^2_{\rm b_{\rm k}})$.  
Here a$_k$ and b$_k$ are the coefficients in the Fourier series, 
and $\sigma_{\rm a_{\rm k}}$ and $\sigma_{\rm b_{\rm k}}$ are the 
uncertainties in the coefficients $a_{\rm k}$ and $b_{\rm k}$, respectively. 
Second, we corrected the powers for the binning using equation 2.19 of \cite{klis89} 
and calculated upper and lower limits to the FPs by using the 
method described in \cite{groth75} \citep[and also in][]{vaughan94}. 
Finally, we normalized the FPs by the total power. 
We show in Figure~\ref{fig:harmpow}, the time evolution of the normalized 
harmonic powers in the first three Fourier harmonics. 
We find that the FPs remain fairly constant in time in all 
investigated energy intervals.

%FIGURE 5
%%%%%%%%%%%%%%%%%%%%%%%%%%%%%%%%%%%

%%%%%%%%%%%%%%%%%%%%%%%%%%%%%% PULSED COUNT RATE  %%%%%%%%%%%%%%%%%%%%%%%%%%%%

\subsection{Pulsed Count Rates}
\label{sect:pulsedrates}

%FIGURE 6
%%%%%%%%%%%%%%%%%%%%%%%%%%%%%%%%%%%%%%%
\begin{figure*}[t]
\vspace{0.0cm}
\begin{center}
\includegraphics[scale=1.0]{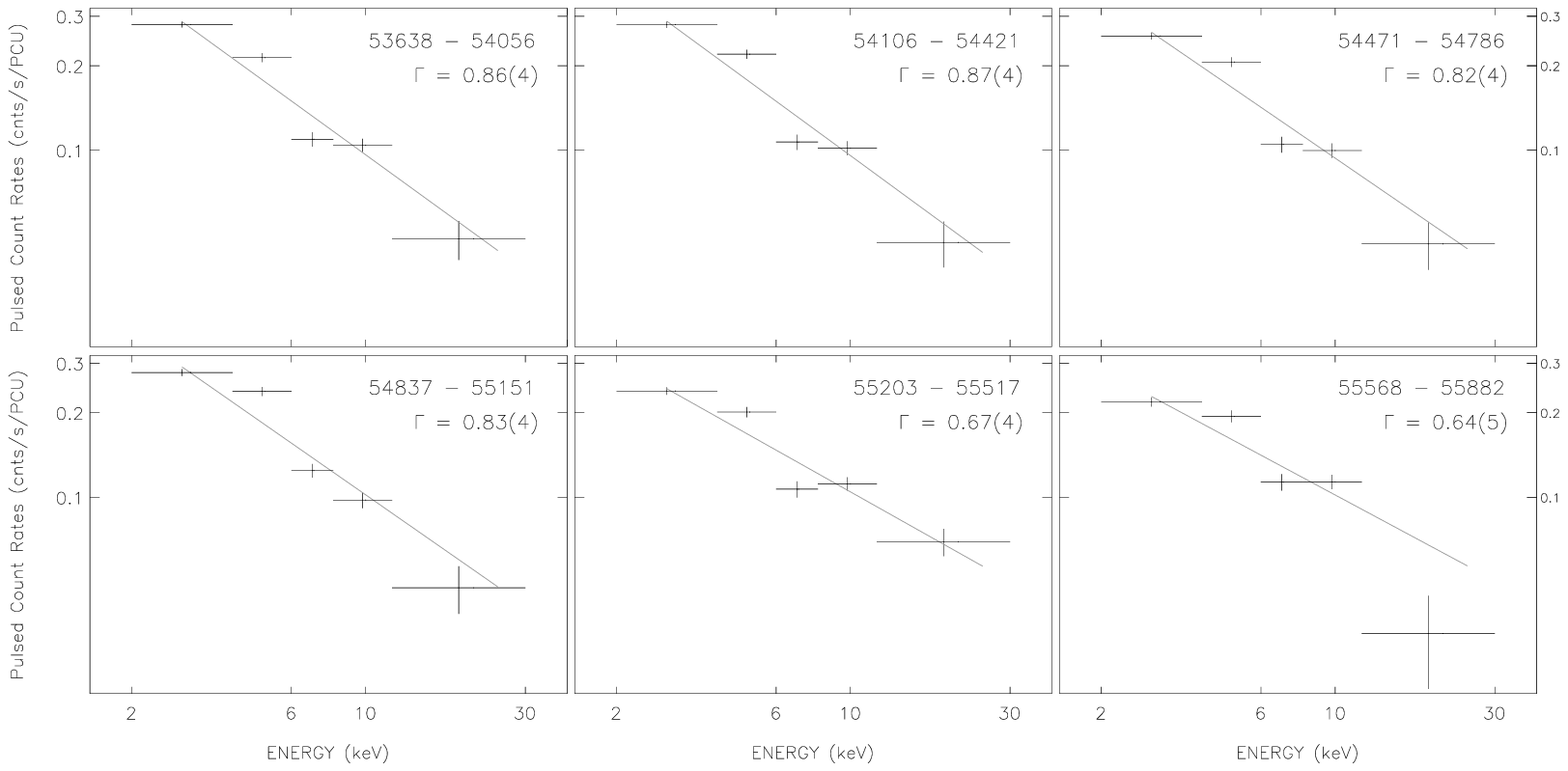}
\caption{Plots of rms pulsed count rates vs. energy. Time intervals 
within which these plots were obtained are shown in the top-right 
of each panel. Solid lines show the best fit power law trends to 
the corresponding energy dependent RMS pulsed count rates. 
Uncertainties in these power law indices refer to the last digit 
as shown in parenthesis in each panel.}
\label{fig:rmsampendept}
\end{center}
\end{figure*}

PCA is not an imaging instrument; it collects all events originating 
within about 1$^{o}$ (FWHM) field centered near the position of \rxsnos. 
Therefore, we cannot construct a precise X-ray light curve of the 
source using PCA observations since the accurate determination of X-ray 
background with the PCA is not possible. Nevertheless, we can trace the 
behavior of the pulsed X-ray emission of \rxs since there is no other 
pulsed X-ray source with exactly the same pulse period in the vicinity. 
X-rays originating from the other sources in the field of view 
(even the pulsed ones) are averaged out after folding the data with 
the spin frequency of \rxs and remain within the DC level. For these 
reasons, we calculated the rms pulsed count rates of the source using

\begin{equation}
{\rm PCR}_{\rm rms} =  \left(\frac{1}{\rm N}\sum_{i=1}^{\rm N}
({\rm R}_{\rm i}-{\rm R}_{\rm ave})^2 -
\Delta {\rm R}_{\rm i}^2\right)^\frac{1}{2}
\end{equation}

\begin{equation}
\delta{\rm PCR_{\rm rms}} = \frac{1}{{\rm N}{\rm PCR}_{\rm rms}}
\left(\sum_{i=1}^{\rm N}
\left[({\rm R}_{\rm i}-{\rm R}_{\rm ave})
\Delta {\rm R}_{\rm i}\right]^2\right)^\frac{1}{2}
\label{eq:rms}
\end{equation}

where ${\rm R}_{\rm i}$ are the count rates in each phase bin, 
$\Delta{\rm R}_{\rm i}$ are their uncertainties, ${\rm R}_{\rm ave}$ 
is their average and N is the number of phase bins. Note that this 
is a background exempt representation of pulsed intensity of the source. 

In Figure~\ref{fig:spindervflux}(d) we present the time variation 
of rms pulsed count rates in the 2$-$10~keV energy range. Here, 
each pulsed intensity value is an average of about 1 month of data 
accumulation. We find that the rms pulsed count rate in the 2$-$10 
keV band does not show any significant variation. 
Figure~\ref{fig:rmsampendept} presents the pulsed count rates as 
a function of energy (in other words, rough energy spectra of the 
pulsed X-ray emission from \rxsnos). Power law fits to these rough 
energy spectra yield a general trend from a more steep shape to a more 
shallow one as time progresses.

%%%%%%%%%%%%%%%%%%%%%%%%%%%%%%     GLITCHES   %%%%%%%%%%%%%%%%%%%%%%%

\subsection{Search for Glitches}
\label{sect:glitches}

\begin{figure}[h!]
\vspace{0.0cm}
\begin{center}
\includegraphics[scale=0.5]{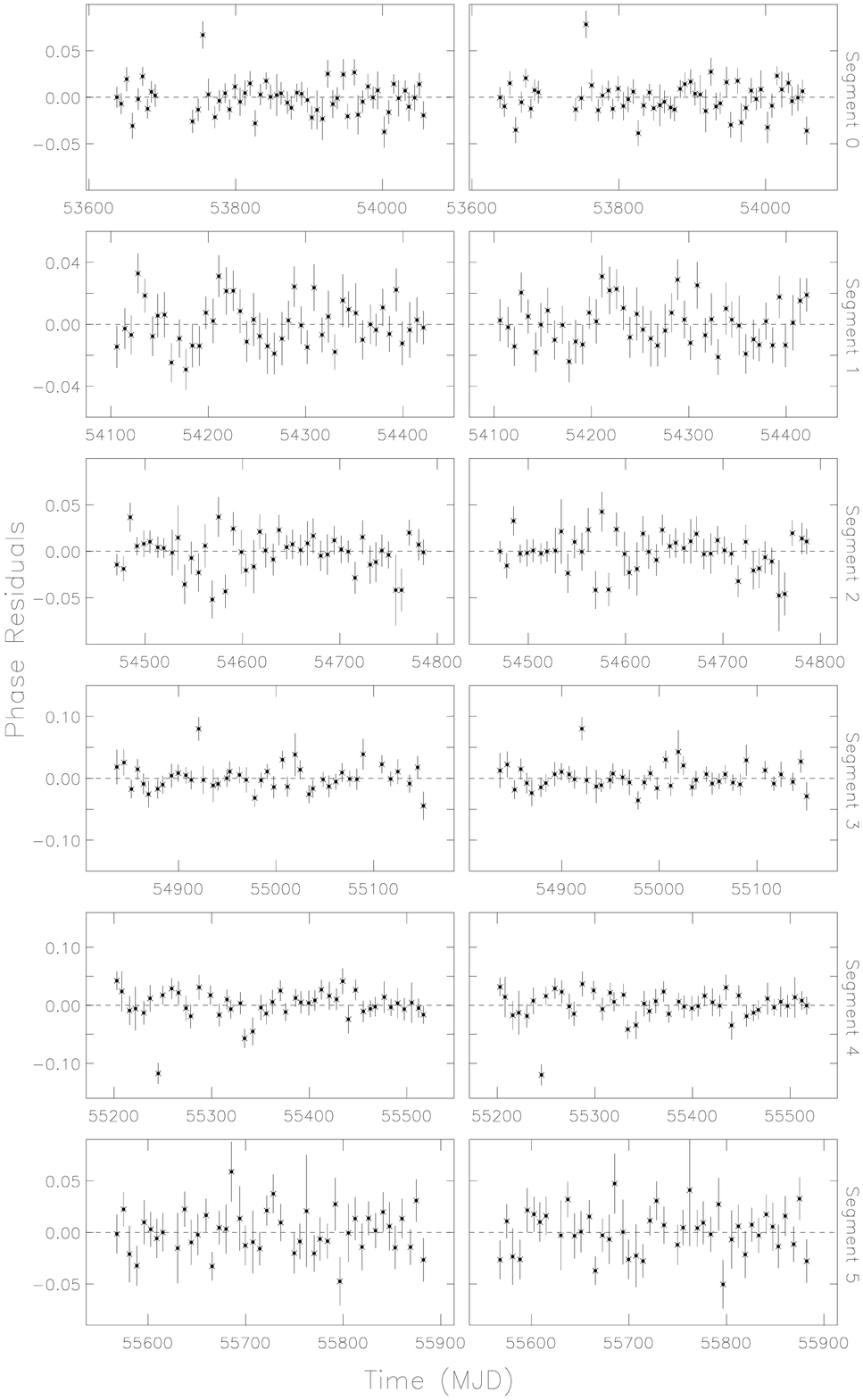}
\caption{({\it Left column}): Phase residuals of the polynomial 
fit to each data segment. ({\it Right column}): Phase residuals 
of the glitch model fit.}
\label{fig:sres}
\end{center}
\end{figure}

There is no explicit glitch detected in our data sample as it can be 
seen from the fit results to the phase drifts in Table 
\ref{tab:tablemain}. To investigate whether there are any small 
amplitude variations in phase drifts (i.e. frequency jumps), we 
fitted phase shifts using the MPFITFUN \footnote{http://purl.com/net/mpfit} 
\citep{markwardt09} procedure which performs Levenberg-Marquardt 
least-squares fit with the corresponding phases of a glitch model 
containing a jump in every $\sim$ 0.1 day and a linear decay, as follows:
\begin{equation}
\nu(t) = \nu_{0}(t) + \Delta{\nu} + \Delta{\dot{\nu}}(t - t_{g})
\label{eq:glt}
\end{equation}
where $\nu_{0}(t)$ is the preglitch frequency evolution, $\Delta{\nu}$ 
is the frequency jump, $\Delta{\dot{\nu}}$ is the change of the 
frequency derivative after the glitch and $t_{g}$ is the epoch of the 
glitch. First we applied this methodology to a previously published 
glitch in 2005 June and a candidate glitch in 2005 September. 
We detected the frequency jumps ($\Delta{\nu}$) and glitch epochs in 
agreement with the published values \citep{israel07a,dib08}. We then 
carried out the glitch search in all six epochs listed in 
Table~\ref{tab:tablemain} as follows: For each epoch, we analyzed 
the fit results on the $\Delta{\nu}$ versus the reduced $\chi^{2}$ plane and 
identified the set of parameters corresponding to the lowest reduced 
$\chi^{2}$ value. We then computed rms fluctuations of phase residuals 
using the possible glitch parameters and compared them with those obtained 
using the polynomial fit results listed in Table~\ref{tab:tablemain}. 
We find that rms phase residual fluctuations with respect to the glitch 
model fits do not indicate any improvement in the fit quality compared 
to the polynomial fits (Figure~\ref{fig:sres}). Moreover, the largest 
glitch amplitude ($\Delta{\nu}$) obtained is about 3$\times$10$^{-8}$ 
Hz in segments 0, 1 and 5 which could well be due to random fluctuations 
of phases, as can be seen in Figure~\ref{fig:sres}.

Consecutive {\it RXTE} observations were typically performed at 7$-$10 
day time intervals. Due to Sun constraints, there were five longer 
gaps of $\sim$50 days in our data set. In order to assess the 
probability for the detection of a glitch that might have 
occurred during these longer gaps, we adopted the detectability 
criterion defined as \citep{alparho83,alpar94}:
\begin{equation}
\delta{\nu} + \delta{\dot{\nu}} * \Delta{t} \ll \Delta{\nu}
\label{eq:crt}
\end{equation}
where $\delta\nu$ and $\delta\dot{\nu}$ specify the total error 
on the spin frequency and frequency derivative determined on both 
ends of the gap, $\Delta$t denotes the duration of the gap, and 
$\Delta\nu$ is the change in spin frequency due to a putative 
glitch. Equation~\ref{eq:crt} implies that $\Delta\nu$ has to be 
much bigger than maximum phase error accumulated across the gap 
in order to identify it as a possible glitch event. We 
calculated the total phase error for each gap adopting the timing 
solutions on both sides of the gaps, and present these results 
in Table~\ref{tab:gapglitch}.

\begin{figure}[h!]
\vspace{0.0cm}
\begin{center}
\includegraphics[scale=0.5]{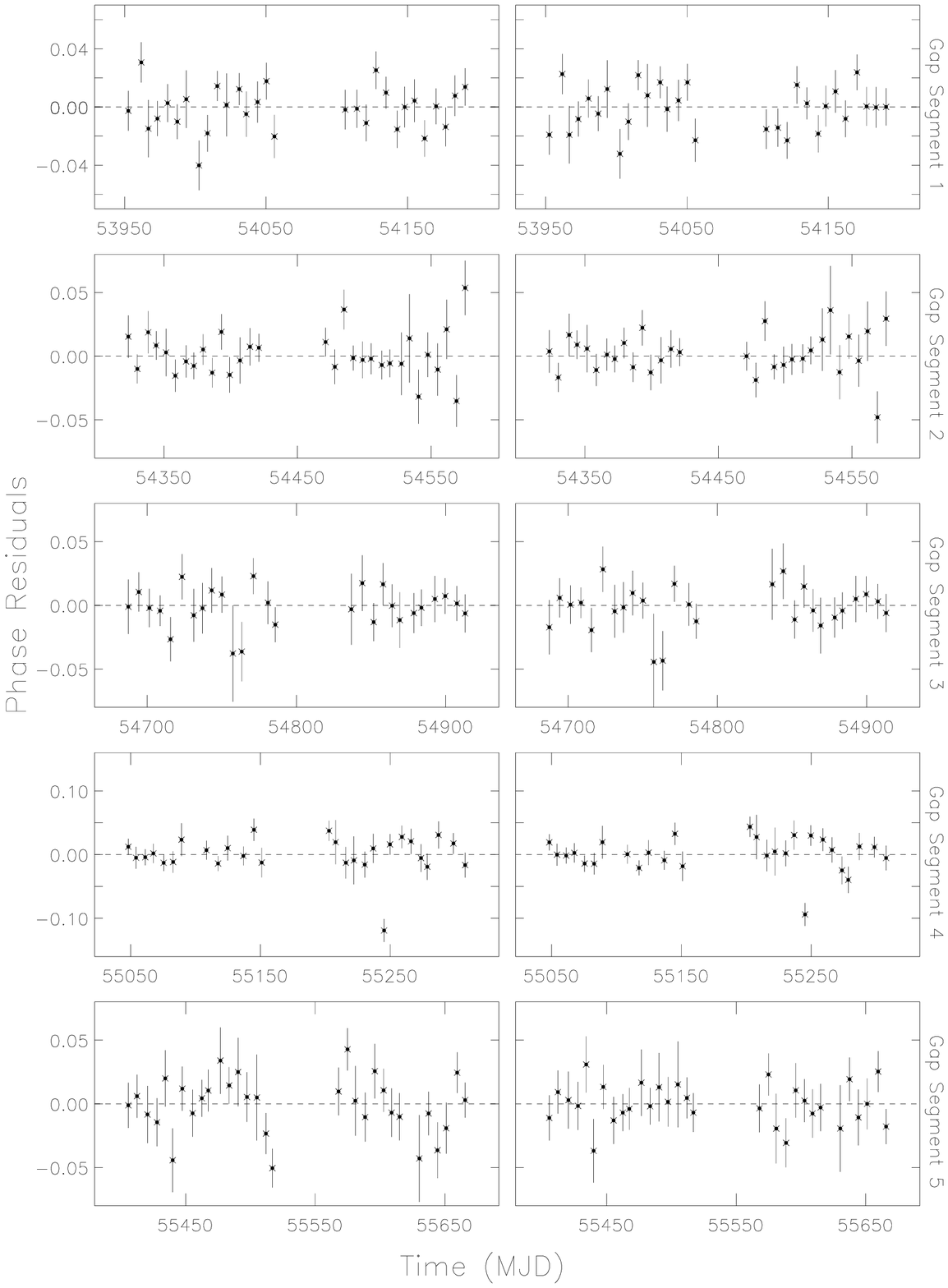}
\caption{({\it Left column}): Phase residuals of the polynomial 
fit to each gap segment. ({\it Right column}): Phase residuals 
of the glitch model fit.}
\label{fig:gapres}
\end{center}
\end{figure}

We then applied the glitch search methodology to $\sim$250 day 
long data segments centered around each gap (gap segment), 
and evaluated minimum $\chi^{2}$ searches as explained above. 
Best-fit timing solutions are listed in Table~\ref{tab:gapglitch}.
Among all gaps, only glitch amplitudes in gap segment 3 
(54687$-$54913) and gap segment 5 (55406$-$55666) satisfy condition 
\ref{eq:crt}. In particular, the glitch amplitude in gap segment 5 
is $\sim$7 times larger than the noise criterion which makes it a 
rather strong candidate for a possible glitch event. The putative 
glitch identified in gap segment 3 has an amplitude $\sim$4 times 
larger than the corresponding minimum noise criterion. The amplitudes 
of estimated glitch events in gap segments 1, 2 and 4 possess large 
errors. The rms fluctuations of phase residuals in gap segments are 
similar in gap segments 1, 2, 3, and 5, while they are much larger in gap 
segment 4. Note that glitch amplitude in gap segment 4 is affected 
from an outlier phase measurement (see Figure~\ref{fig:gapres}), 
without which the glitch amplitude becomes even less significant. 
We, therefore, identified two glitch candidates; a strong case in 
the gap segment 5, and another one in gap segment 3 which is 
slightly less robust. We discuss their implications below.

%TABLE 2
%%%%%%%%%%%%%%%%%%%%%%%%%%%%%%%%%%%%%%%%%%%%%%%%%%%%%%%
\begin{deluxetable*}{lccccc}
\tablewidth{0pt}
\tabletypesize{\scriptsize}
%\rotate
\tablecaption{Timing Solutions in the Segments Including the Gaps
\label{tab:gapglitch}}
\tablehead{
\colhead{Parameter}&
\colhead{Gap Segment 1}&
\colhead{Gap Segment 2}&
\colhead{Gap Segment 3}&
\colhead{Gap Segment 4}&
\colhead{Gap Segment 5}
}
\startdata 
Range (MJD)&53952.558$-$54190.841&54323.309$-$54575.557&54687.314$-$54913.097&55048.230$-$55307.736&55405.757$-$55665.660\\
Epoch (MJD)&54106.040&54471.050&54836.804&55202.849&55567.977\\
Number of TOAs&29&31&27&28&31\\
$\nu$ (Hz)&0.0908775775(8)&0.090872564(1)&0.090867525(2)&0.090862442(2)&0.090857270(4)\\
$\dot{\nu}$ ($10^{-13}$ Hz s$^{-1}$)&$-$1.617(2)&$-$1.584(2)&$-$1.590(4)&$-$1.646(3)&$-$1.638(4)\\
$t_{g}$ (MJD)& 54174.105&54531.016&54818.531&55245.277&55532.328\\
$\Delta{\nu}$ ($10^{-8}$ Hz)&8(4)&2(2)&6.4(4)&4(1)&12.4(3)\\
$\Delta{\dot{\nu}}$ ($10^{-15}$ Hz s$^{-1}$)&$-$43(56)&9(9)&$-$3(1)&$-$10(4)&$-$4.3(8)\\
rms (phase) &0.0151&0.0165&0.0170&0.0264&0.0156 \\
\hline
Gap Range (MJD) & 54056$-$54106 & 54422$-$54471 & 54786$-$54836 & 55151$-$55202 & 55517$-$55567\\
Gap Criterion ($10^{-8}$) & 2.31 & 2.15 & 1.76 & 0.73 & 1.74
\enddata
\tablenotetext{a}{Values in parenthesis are the uncertainties in the 
last digits of their associated measurements}
\end{deluxetable*}

%FIGURE 7
%%%%%%%%%%%%%%%%%%%%%%%%%%%%%%%%%%%%%%%%

%FIGURE 8
%%%%%%%%%%%%%%%%%%%%%%%%%%%%%%%%%%%%%%%%%

%TABLE 3
%%%%%%%%%%%%%%%%%%%%%%%%%%%%%%%%%%%%%%%%%%%%%%%%%%%%
\begin{deluxetable*}{cccccc}
\tablewidth{0pt}
\tabletypesize{\scriptsize}
%\rotate
\tablecaption{Critical parameter values and results of the 
expectancy analysis of \rxs
\label{tab:gltexp}}
\tablehead{
\colhead{Number of glitches}&
\colhead{Critical parameter\tablenotemark{a}}&
\colhead{}&
\colhead{Expectancy of Glitches\tablenotemark{b}}&
\colhead{}\\
\colhead{(n)}&
\colhead{($\langle\delta\Omega/\Omega\rangle$)}&
\colhead{1998$-$2005}&
\colhead{1998$-$2011}&
\colhead{2006$-$2011}
}
\startdata       
   &2.3$\times$10$^{-4}$&     1.8 &          3.2       & 1.4\\
2  &2.0$\times$10$^{-4}$&     2.0 &          3.6       & 1.6\\
   &1.6$\times$10$^{-4}$&     2.5 &          4.6       & 2.0\\
\hline   
   &1.5$\times$10$^{-4}$&     2.6 &          4.8       & 2.1\\ 
3  &1.4$\times$10$^{-4}$&     3.0 &          5.4       & 2.4\\
   &1.1$\times$10$^{-4}$&     3.8 &          6.9       & 3.0\\  
\hline  
   &7.7$\times$10$^{-5}$&     5.3 &          9.2       & 3.8\\
6  &6.8$\times$10$^{-5}$&     6.0 &          10.4      & 4.2\\
   &5.4$\times$10$^{-5}$&     7.6 &          13.2      & 5.4
\enddata
\tablenotetext{a}{The upper (top), average (middle) and lower 
values (bottom) for the critical parameter value of the vortex 
unpinning model.}
\tablenotetext{b}{Calculated using the average value of the 
$\dot{\nu}$/$\nu$ within the specified time range.
Timing solutions before 2005 are taken from \cite{dib08}.}
\end{deluxetable*}

%%%%%%%%%%%%%%%%%%%%%%%%%%%%%%%%%%%%%%%%%%%%%%%%%%%%%%%%%%%%%%%%%%%%%%%%%
%DISCUSSION
%%%%%%%%%%%%%%%%%%%%%%%%%%%%%%%%%%%%%%%%%%%%%%%%%%%%%%%%%%%%%%%%%%%%%%%%%

\section{Discussion and Conclusions}
\label{sect:discuss}

We performed detailed long term timing studies of \rxs spanning 
$\sim$6 yr. Together with the earlier extensive study of the 
source by \cite{dib08}, our investigation considers the entire 
database of {\it RXTE} observations of \rxsnos. In our long-term 
timing investigations, it was possible to describe the phase 
shifts with a second order polynomial in only one interval 
(Segment 4 in Table~\ref{tab:tablemain}), while all other parts 
required higher order terms. These results are similar to what 
has been obtained by \cite{dib08}, \cite{archibald08}, and 
\cite{israel07a}, confirming the fact that \rxs is indeed a 
noisy pulsar.

The pulse profile of \rxs in the 2$-$10 keV band does not show 
any significant variations over the last $\sim$6 yr, maintaining 
its general pulse structure as in the earlier epochs. A minor 
structure (described as a shoulder above) in the pulse profile 
below 4 keV becomes stronger with energy and dominates the pulse 
profiles above 8 keV, as also noted by \cite{hartog08} regarding 
earlier observations of the source. We also find no significant 
changes in the rms pulsed count rates (i.e., a measure of the 
pulsed flux) in the 2$-$10 keV range. In these respects, \rxs 
exhibits an almost stable pulsed X-ray emission behavior. We 
constructed a coarse energy spectrum of the rms pulsed count 
rates for each observation segment and found that it becomes 
gradually harder with time, as indicated by a shallowing power 
law index. 

As a result of $\sim$14 yr of {\it RXTE} observations, three 
glitches with two different recovery characteristics were 
unveiled unambiguously, and three candidate glitches were suggested 
in the time baseline between 1999 and 2005. Such a glitching 
behavior of \rxs made this system one of the most frequently 
glitching pulsars \citep{israel07a,dosso03,dib08}. It is important 
to report the fact that, we do not find any unambiguous glitches 
in the time interval between 2006 and 2011. However, glitch 
search in the gaps yielded a strong candidate in gap 5 with glitch 
amplitude $\sim$$10^{-7}$ which is $\sim$7 times larger than the 
noise in this gap and on the order of largest glitches observed 
from this source. We identified another candidate in gap segment 3, 
although it is slightly less robust.
 
Glitches are generally explained by models involving the neutron 
star crust, superfluid component of the inner crust or core superfluid 
and starquakes. The superfluid vortex unpinning model involves the 
crust and inner crust superfluid \citep{anderson75,alpar84a}. 
In this model vortices formed by superfluid are pinned to the 
neutron-rich nuclei. While the crust spins down due to the 
electromagnetic torques, a rotational lag between the superfluid 
component and the crust builds up. When a critical value of 
rotational lag ($\delta\Omega$ $\equiv$ $\Omega_{s}$ - $\Omega_{c}$, 
where $\Omega_{s}$ and $\Omega_{c}$ are the superfluid's and 
crust's rotational rate, respectively) is reached, vortices suddenly 
unpin, resulting in transfer of angular momentum to the crust, i.e., 
glitch. This lag also determines the glitch occurrence time interval. 
This model is successful in explaining large glitches 
($\Delta\nu$/$\nu$ $\sim$ 10$^{-6}$), such as those observed from 
the Vela pulsar with an occurrence time interval of $\sim$2 yr 
\citep{alpar81,alpar84b}.

Another class of models invokes starquakes, which are triggered by the 
cracking of the solid neutron star when growing internal 
stresses strain the crust beyond its yield point 
\citep{ruderman69,ruderman76,ruderman91,baym71}. This critical 
strain can be reached due to several mechanisms: the star spin 
down causes a progressive decrease of the equilibrium oblateness of 
the crust \citep{ruderman69,ruderman91,franco00}; variations 
of the core magnetic field, due to the motion of core superfluid 
vortices coupled to it \citep{srinivasan90, ruderman98}; and, 
in strongly magnetized neutron stars, the rapid diffusion of the 
core magnetic field (or the "turbulent" evolution of the crustal 
field) provides an alternative channel to produce crustal fractures 
\citep{thompson96, rheingepp02}. \cite{dosso03}, based on the 
different recovery characteristics of the glitches of \rxsnos, 
proposed that they can be explained by a magnetically-driven 
starquake model since they intrinsically involve local processes 
and a higher degree of complexity.

In order to discriminate between different possible models, 
\cite{alpar94} following \cite{alparho83}, investigated 
the global properties of large pulsar glitches using a sample 
of 430 pulsars, excluding the Vela pulsar. As these sources 
are not continuously monitored due to limited telescope times 
or other observational constraints, there are unavoidable data 
gaps in between successive pointings. This case puts a serious 
constraint on the detectability of a glitch if it occurs in a 
data gap of a pulsar with noisy timing behavior. They introduced 
a noise criterion (see Eqn. \ref{eq:crt}) for significantly 
detecting frequency jumps in the observational gaps. Therefore, 
they restricted their analysis to the 19 pulsar glitches with 
$\Delta{\nu}$/$\nu$ $>$ $10^{-7}$. They estimated the physical 
parameters, e.g., inter-glitch time for the vortex unpinning 
model and the glitch size for the core-quake model. The parameters 
of the former model were estimated with two different assumptions 
for unpinning: First, the critical glitch parameter is taken as 
$\delta\Omega$ which is a representative of the number of vortices 
that is unpinned at the time of the glitch. Second, this parameter
is taken  as fractional density of the unpinned vortices that is 
proportional to $\delta\Omega/\Omega$, as the density of vortices 
$\propto$ $\Omega$. They also assumed that the probability of 
observing $n$ glitches is given by Poisson statistics. Glitch size 
estimation from the core-quake model is far bigger than the glitch 
amplitudes of the Vela pulsar and sample mean. Thus, their work 
statistically excluded the core-quake model. They also compared 
the parameter estimates of the vortex unpinning model with those 
of glitches from Vela and other pulsars, and concluded that the 
vortex unpinning model with a constant fractional vortex density 
($\langle\delta\Omega/\Omega\rangle$) is the most compatible model 
and can represent an invariant for glitches.

To test the glitch expectancy within the vortex unpinning model for 
\rxs glitches, we applied the same statistical glitch expectancy 
analysis \cite[see Equation 11 of][]{alpar94} and estimated the 
expected number of glitches using $\sim$14 yr of {\it RXTE} 
observations. We calculated the critical fractional vortex 
density of the vortex unpinning model by using the time span 
between 1998 January and 2005 November, which contains three 
glitches and three glitch candidates \citep{dib08,dosso03,israel07a}. 
For a single pulsar, $\dot{\nu}$/$\nu$ value is not expected to 
fluctuate between observations. However, this is not the case for 
\rxs as it changes between -1.87$\times$10$^{-12}$ s$^{-1}$ and 
-1.31$\times$10$^{-12}$ s$^{-1}$ with an average value of 
-1.66$\times$10$^{-12}$ s$^{-1}$, which further implies the noisy 
timing characteristics of the source. Therefore, we performed our 
calculations for all these three values. First we included the 
observational gaps into the total time span which, by the chosen
noise criterion, restricts our analysis to large glitches with 
$\Delta{\nu}$/$\nu$ on the order of $10^{-6}$. Using 
$\dot{\nu}$/$\nu$ values and observed number of glitches with 
$\Delta{\nu}$/$\nu$$\sim$10$^{-6}$ (i.e., n = 2), we obtain the 
upper, lower and average values for critical parameter value of 
the vortex unpinning model. We note an important fact here that 
a glitch candidate (i.e., near candidate glitch 2 in \cite{dib08}) 
was reported by \cite{israel07a} with a fractional amplitude 
of 1.2$\times$10$^{-6}$. If the latter report is correct, the 
number of large glitches in the 1998$-$2005 interval would be 3 
(i.e, n = 3) which changes the critical parameter. Finally, we 
excluded all data gaps except the ones with glitches reported 
in them, and the ones that satisfied the noise criterion in our 
analysis, and we considered all reported glitches with 
$\Delta{\nu}$/$\nu$ $\gtrsim$ 10$^{-7}$ (i.e., n = 6) and 
calculated the critical parameters for this case as well. In 
Table~\ref{tab:gltexp} we list the values of the critical 
parameter for each of the above-mentioned cases and their 
corresponding expected number of glitches in the time intervals 
between 1998$-$2005, 1998$-$2011, and 2006$-$2011. As expected, 
the average value of the critical parameter yields the observed 
number of glitches in the 1998$-$2005 interval. We find that the 
total number of expected glitches with fractional amplitudes 
of $\gtrsim$$10^{-6}$ (n = 2 in Table~\ref{tab:gltexp}) varies 
between 3.2 and 4.6 if the time baseline spans untill the end of 
the {\it RXTE} coverage of the source in 2011 November. The number 
of glitches in the 2006$-$2011 time range, where we found a strong 
candidate, were expected to range from 1.4 to 2.0. We then 
repeated the above procedure, this time excluding all data gaps 
except the ones with reported candidate glitches. In this case, the 
noise criterion allows consideration of all glitches with 
$\Delta{\nu}$/$\nu$ $\sim$10$^{-7}$ (i.e., n = 6), and we 
re-calculated the critical parameters (see Table~\ref{tab:gltexp}).

Glitch expectancy analysis within the context of vortex unpinning 
model suggests that \rxs might have had, on average, two large glitches 
in 6 yr, corresponding to the interval of 2006$-$2011 
(Table~\ref{tab:gltexp}). The two significant glitch candidates 
we identified in gap segments are, therefore, important, since 
they comprise the observed number to match with the expectancy of the 
vortex creep model. As far as only glitch statistics is concerned, 
this case implies that the mechanism leading to the observed glitches 
in \rxs is internal. However, where particular glitch characteristics 
were concerned (e.g., discrepancies in glitch recovery), the vortex 
unpinning model is argued to be not sufficient \citep{dosso03}. 

\rxs is the only member of the magnetar family that has not 
exhibited energetic X-ray bursts. Almost all other AXPs, that 
have experienced timing glitches, emitted energetic bursts 
either in conjunction with (e.g., \ctb2259, \cite{woods04}) 
or contemporaneous to their glitches. It is, therefore, 
suggestive that a common mechanism might be responsible for 
both glitches and bursts. The dipole magnetic field strength 
of \rxs as inferred from its spin period and spin-down rate is 
about 4.6$\times$10$^{14}$ G, that is strong enough to produce 
significant deformation in the neutron star crust and eventually 
lead to the release of energy via bursts \citep{thompson95}. 
Nevertheless, the condition on \rxs has not given rise to any 
observable bursts, even though it has experienced the largest 
number of glitches among all magnetars. While a common 
mechanism could reproduce coincident energetic bursts and glitches 
in general, it might be generating glitches but not detectable 
enhancements and bursts in \rxsnos, possibly due to this source 
having slightly lower crust shear modulus, so that the release of 
less energy can still produce breaks in the crust. The energetic 
bursts, however, are not accounted for within the context of the 
vortex unpinning model which appears to be favored for this source
in our statistical investigations.

Recently \cite{eichler10} suggested that vortices can be unpinned 
mechanically via oscillations rather than by a sudden heat release. 
According to their estimation, the relative velocity between the 
crust and superfluid, which is generated by the mechanical energy 
release at the depths below 100 m, can exceed the critical velocity 
lag and unpin the vortices. In order to explain the radiatively 
silent glitches seen in some AXPs (as in the case of \rxsnos) they 
proposed that mechanically triggered glitch event might not be 
accompanied by a long-term X-ray brightening since a glitch can be 
triggered by a less energy release. In this picture, the origin of 
X-ray brightening is also through mechanical energy release and 
these flux enhancements are expected to be accompanied with glitch 
events. This scenario can be diagnosed through the exact timing of 
glitches with radiative enhancements  \citep{eichler10}.

%%%%%%%%%%%%%%%%%%%%%%%%%%%%%%%%%%%%%%%%%%%%%%%%%%%%%

\acknowledgments

We would like to thank M. Ali Alpar and the anonymous referee for 
helpful comments. S\c{S}M acknowledges support through the national 
graduate fellowship program of the Scientific and Technological 
Research Council of Turkey (T\"UB\.{I}TAK).

\end{document}